\documentclass[5p]{elsarticle} 

\journal{N/A}

\biboptions{numbers,sort&compress}

\usepackage{eurosym}
\usepackage{url}
\usepackage{amsmath}
\usepackage{graphicx}
\usepackage[colorlinks]{hyperref}

\usepackage[
	type={CC},
	modifier={by},
	version={4.0},
]{doclicense}

\usepackage{lineno}
\modulolinenumbers[1]

\begin{document}

\begin{frontmatter}

\title{Costs of Regional Equity and Autarky in a Renewable European Power System}

\author[kitaddress]{Fabian Neumann\corref{correspondingauthor}}
\ead{fabian.neumann@kit.edu}
\cortext[correspondingauthor]{Corresponding author}
\address[kitaddress]{Institute for Automation and Applied Informatics (IAI), Karlsruhe Institute of Technology (KIT), Hermann-von-Helmholtz-Platz 1, 76344, Eggenstein-Leopoldshafen, Germany}

\begin{abstract}
Social acceptance is a multifaceted consideration when planning
future energy systems, yet often challenging to address endogeneously.
One key aspect regards the spatial distribution of investments.
Here, I evaluate the cost impact and changes in optimal system composition when development
of infrastructure is more evenly shared among countries
and regions in a fully renewable European power system.
I deliberately deviate from the resource-induced cost optimum
towards more equitable and self-sufficient solutions in terms of power generation.
The analysis employs the open optimisation model PyPSA-Eur.
I show that cost optimal solutions lead to very inhomogenous distributions of assets,
but more uniform expansion plans can be achieved on a national level
at little additional expense below 4\%. Yet completely autarkic solutions, without
power transmission, appear much more costly.
\end{abstract}


\end{frontmatter}


\section{Introduction}
\label{sec:introduction}

\begin{figure}
	\centering
	\scriptsize
	(i) Imbalances in 2018 according to ENTSO-E \cite{entsoe_factsheet_2018}
	\includegraphics[width=0.9\columnwidth, trim=0 0.65cm 0 0, clip]{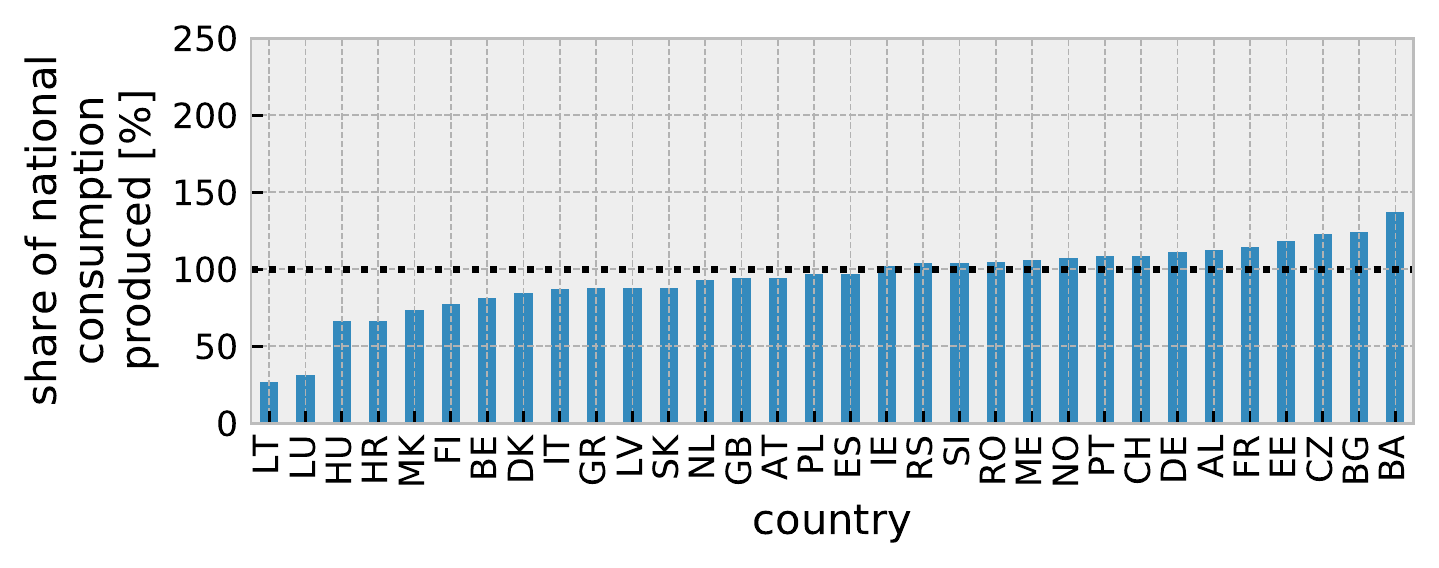} \\
	(ii) Imbalances in optimised fully renewable system \\
	(GR=308\%, NL=360\%, DK=874\%)
	\includegraphics[width=0.9\columnwidth, trim=0 0.65cm 0 0, clip]{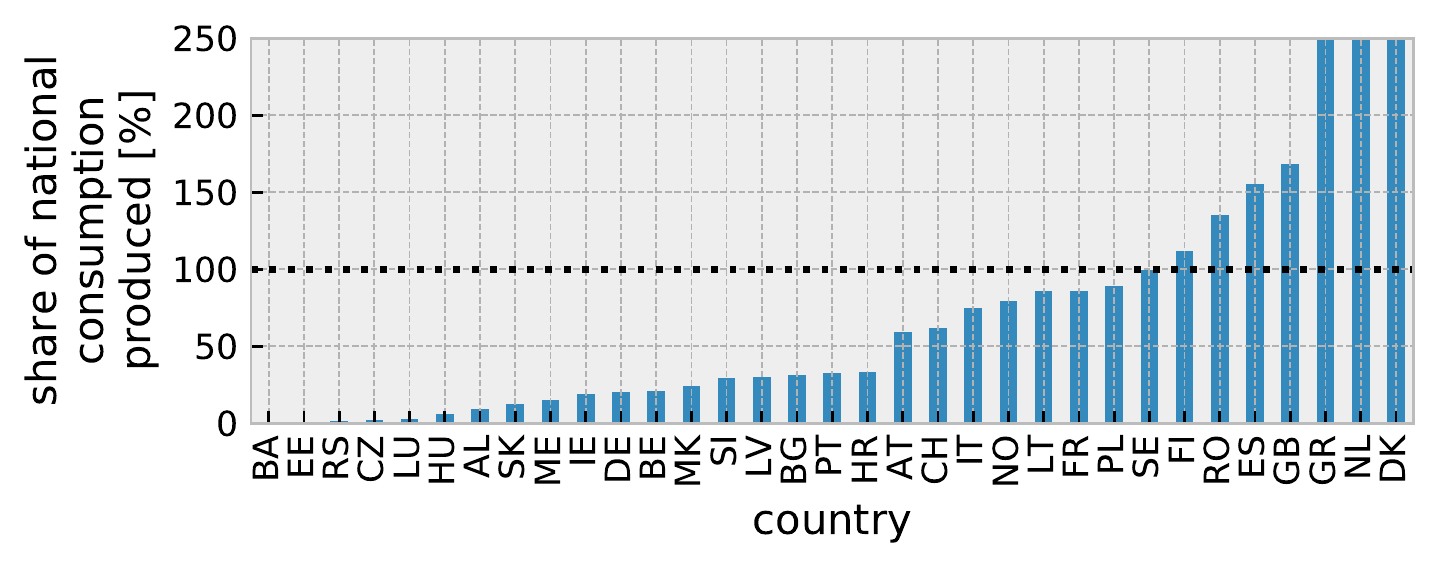}
	\caption{Imbalances today and in optimised renewable system.}
	\label{fig:imbalance}
	\vspace{-0.3cm}
\end{figure}

Optimising for a renewable power system was observed to entail a
very heterogenous distribution of electricity generation in relation
to demand when compared to national imbalances reported by ENTSO-E
for 2018 (Figure \ref{fig:imbalance}) \cite{entsoe_factsheet_2018}.
The system is dominated by many distinct net importers and exporters, whereas
few supply just their own demand.
This raises concerns about distributional equity.

Blindly following the cost optimum risks inequitable outcomes and public headwind,
bearing the potential of decelerating the energy transition.
Particularly wind farms and transmission lines spark local opposition,
which was found to be best counteracted by including the public in the planning process
and by sharing profits \cite{cohen_refocussing_2014}.
Vice versa, also the absence of investments may have a
detrimental impact on local communities.

Beyond the spatial distribution of generation investments,
numerous other dimensions of energy justice and equity principles exist \cite{sasse_distributional_2019}.
Equity metrics can also relate to temporal, income,
racial, labor and environmental aspects \cite{jenkins_energy_2016,mayfield_quantifying_2019, fell_capturing_2020},
and perceptions of fairness vary among stakeholders \cite{batel_social_2013,lehmann_managing_2020}.
Recent developments of pan-continental models with growing sub-national detail raise the need
for recognising their regional implications \cite{li_regional_2016, fell_capturing_2020}.
However, aspects of fairness are challenging to assess in endogenous modelling and
analyses have so far been limited to ex-post analysis \cite{fell_capturing_2020}.
Enhanced collaboration between social scientists and energy
modellers has been encouraged \cite{trutnevyte_societal_2019}.

Moreover, there is a trend towards discussing energy autarky,
i.e.~the ability to operate regions partially or completely independently
\cite{schmidt_regional_2012,engelken_transforming_2016}.
Positive associations with autonomy, control and independence
drive aspirations for self-sufficiency for individuals and
municipalities alike, and result in a higher willingness to pay
and greater support for projects
\cite{muller_energy_2011,ecker_promoting_2017, rae_energy_2012,engelken_transforming_2016}.
The debate also evolves around the resilience of more decentralised systems \cite{panteli_grid_2015}.
The primary resource-based feasibility of autarkic systems on different spatial
levels was evaluated in Tr\"{o}ndle et al.~\cite{trondle_homemade_2019}.
High population density was found to sometimes be a limiting factor for small autarkic systems.
Weinand et al.~found that about half of the 6000 municipalities in Germany
have sufficient potentials to become off-grid municipalities \cite{weinand_identification_2020}.

Further related work has assessed the benefit of transmission capacities
between countries \cite{schlachtberger_benefits_2017} and more
heterogeneous distributions of generation assets \cite{eriksen_optimal_2017}.
It has moreover been evaluated what range of similarly costly
but possibly more socially acceptable
power systems can be realised \cite{neumann_near_2019}
and what costs are incurred by reducing eligible potentials
\cite{bolwig_climatefriendly_2020}. 
Previous work on distributional equity regarding
power generation has however
covered only a single country and neglected
the variability of renewable generation and demand, as well
as the interaction between storage and transmission infrastructure
\cite{drechsler_efficient_2017,sasse_distributional_2019}.
A cost assessment of regional autarkic systems in Europe 
compared to the least-cost system does not appear to exist yet.

In this contribution, I remedy the concerns about
spatial scope and temporal resolution and explore at what cost more
evenly distributed, or even autarkic, power supply could be achieved in Europe,
regarding both countries and smaller regions.

\begin{figure}
	\begin{center}
		\begin{scriptsize}
		  \setlength\tabcolsep{0pt}
		  \begin{tabular}{cc}
			(i) National Requirements & (ii) Nodal Requirements \\
			\includegraphics[width=0.49\columnwidth]{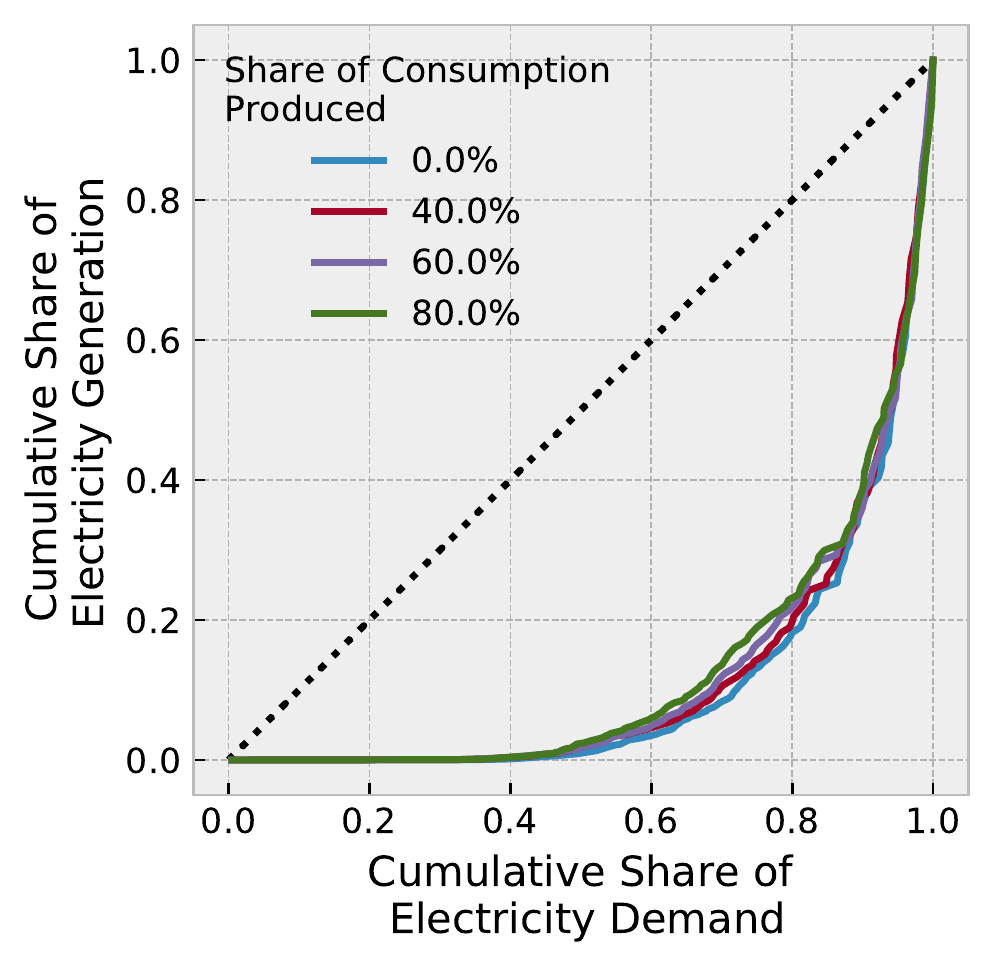} &
			\includegraphics[width=0.49\columnwidth]{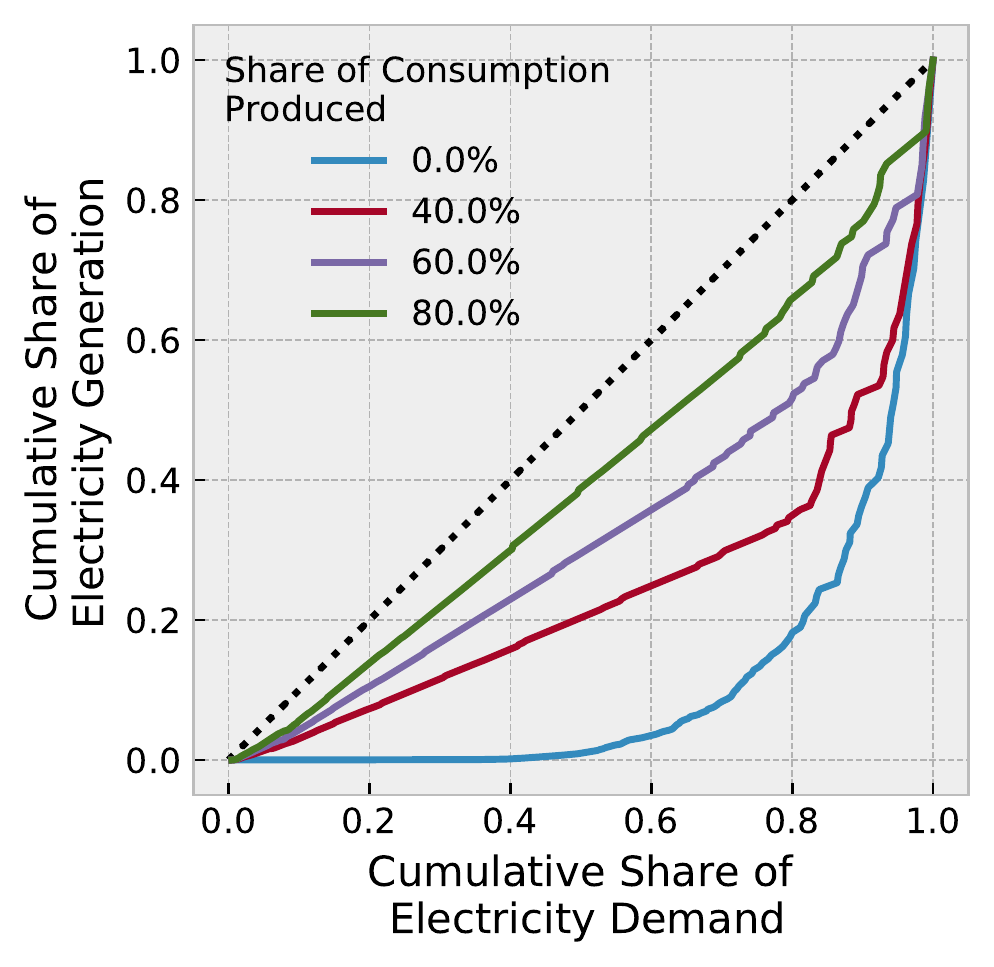}
		  \end{tabular}
		\end{scriptsize}
	  \end{center}
	\vspace{-0.7cm}
	\caption{Lorenz curves for different equity requirements relating the cumulative share of electricity generation to the cumulative share of demand in the 200 regions of the European power system model.}
	\label{fig:lorenz}
	\vspace{-0.3cm}
\end{figure}

\section{Model Setup}
\label{sec:model}

I use the open European transmission system model PyPSA-Eur
with 200 nodes and 4380 snapshots,
one for every two hours in a year \cite{Horsch2018}.
I solve a long-term power system planning problem which seeks to
minimise the total annual system costs comprising generation,
transmission and storage infrastructure in a fully renewable system. 
The objective is subject to linear constraints that define limits on
(i) the capacities of infrastructure from geographical and technical potentials,
(ii) the availability of variable renewable energy sources for each
location and point in time derived from reanalysis weather data, and
(iii) linearised multi-period optimal power flow (LOPF) constraints
including storage consistency equations.

I add constraints for each country or node to produce on
average at least a given share of their annual consumption;
i.e.~I explore the sensitivity of increasing production equity requirements.
The extreme cases are
(i) every country or node produces as much as required for the cost-optimal system
using the most productive locations (0\%) and
(ii) every country or node produces as much as they consume (100\%).
The experiments interpolate between the extremes in steps of 10\%.

I further extend this setup by two experiments regarding absolute autarky:
(i) one where there is no cross-border transmission of power between countries
    but which includes the intranational transmission grid, and 
(ii) one where each node fully supplies its own power demand at any time in isolation.
The code to reproduce all results is available at
\href{https://github.com/fneum/equity-and-autarky}{github.com/fneum/equity-and-autarky}.

\section{Results and Discussion}
\label{sec:results}

\begin{figure}
	\centering
	\scriptsize
	\includegraphics[width=0.9\columnwidth]{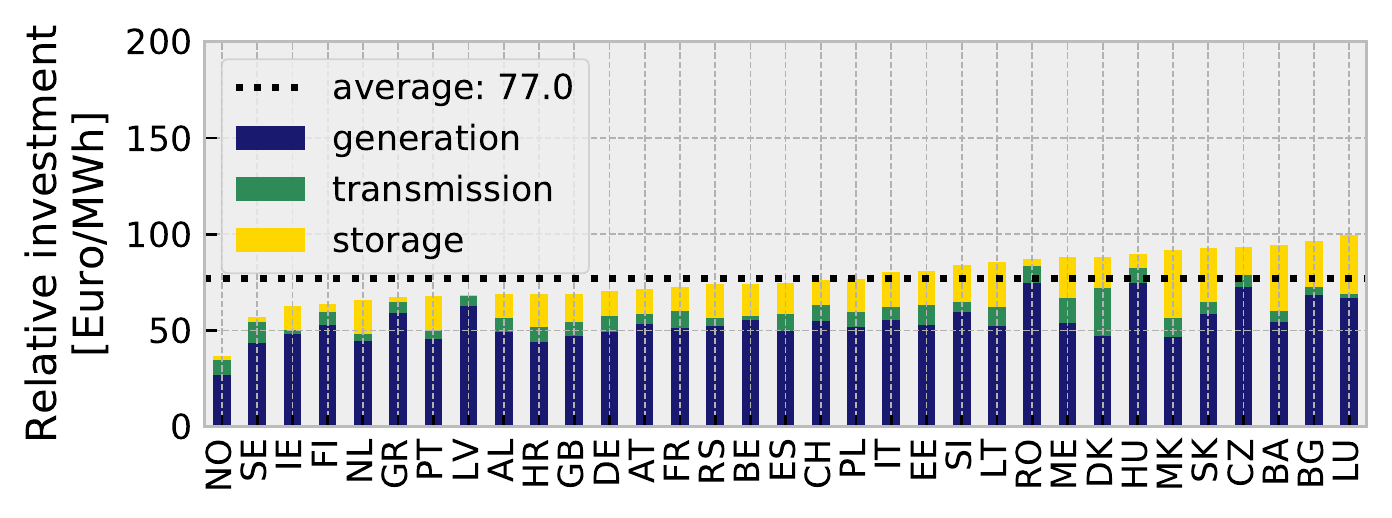} \\
	\vspace{-0.3cm}
	\caption{National annual investment relative to annual demand when every
			 country produces as much as they consume (100\%).}
	\label{fig:invest}
	\vspace{-0.3cm}
\end{figure}

\begin{figure}
	\centering
	\scriptsize
	\includegraphics[width=\columnwidth]{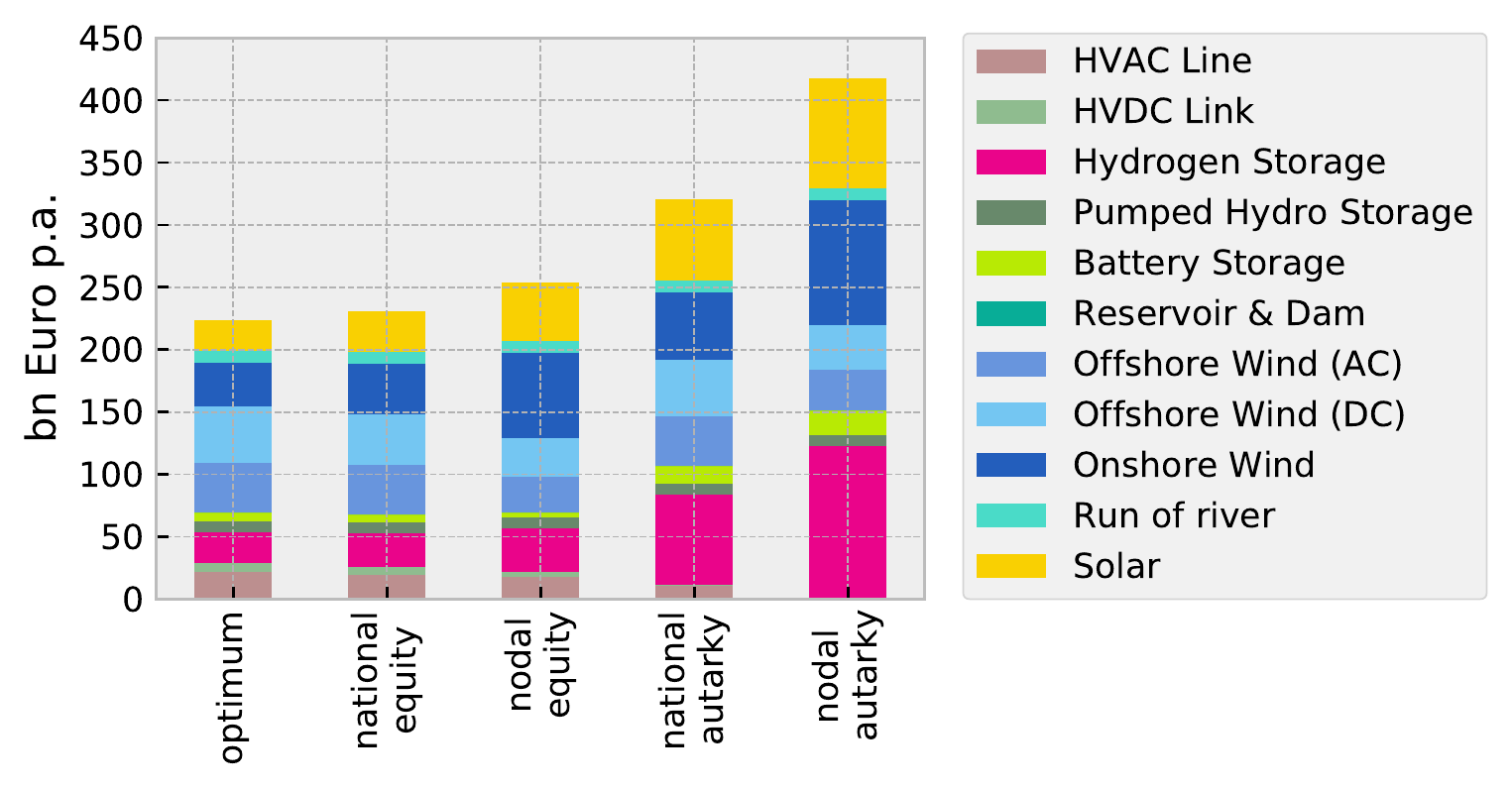} \\
	\vspace{-0.35cm}
	\caption{Total system cost impact of autarky on national and nodal levels
	compared to optimal solution and maximal equity constraints.}
	\label{fig:autarky-cost}
	\vspace{-0.3cm}
\end{figure}


The discussion of results employs
system costs, the technology mix,
as well as the distribution of 
power system infrastructure investments
as evaluation criteria,
both regarding equity and autarky considerations.

\begin{figure*}
	\centering
	\vspace{-1cm}
	\includegraphics[width=0.36\textwidth, trim=0 0 5.9cm 0, clip]{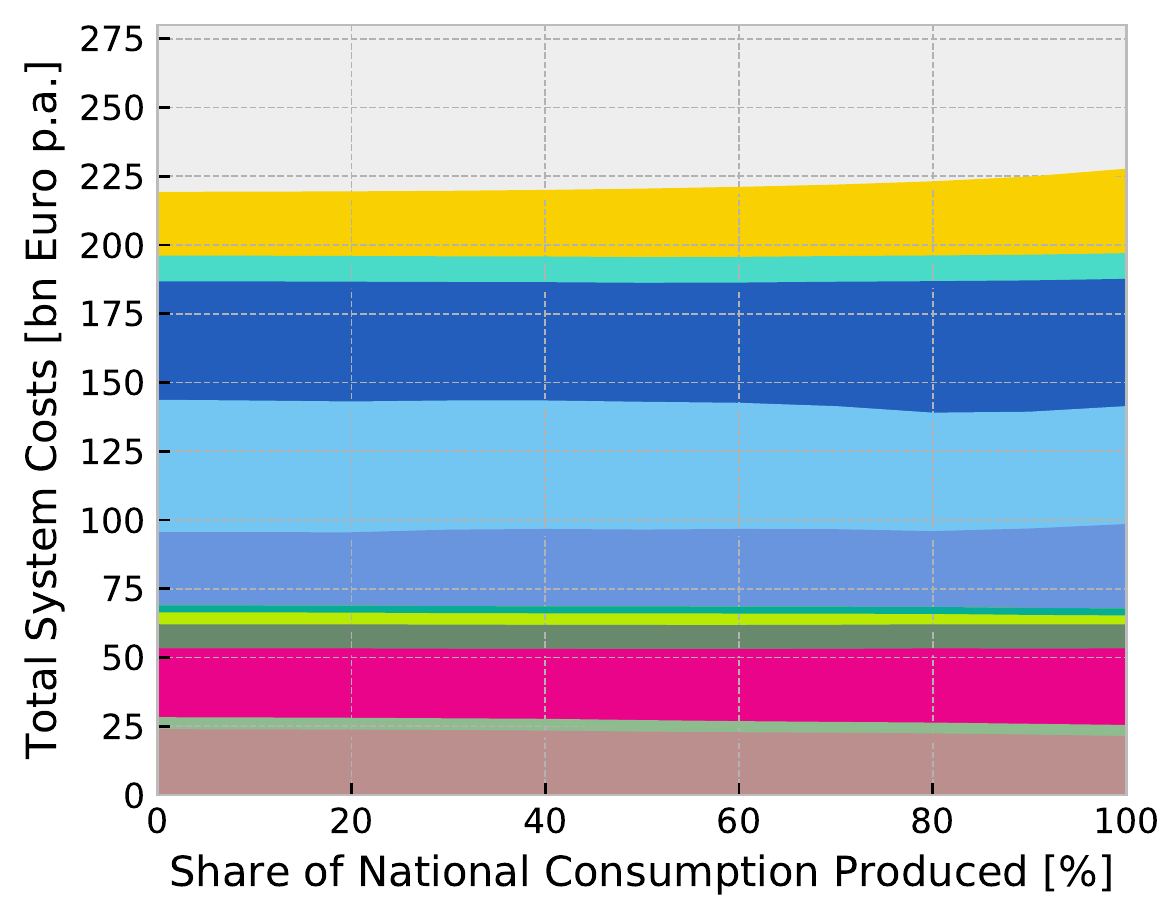}
	\includegraphics[width=0.36\textwidth, trim=0 0 5.9cm 0, clip]{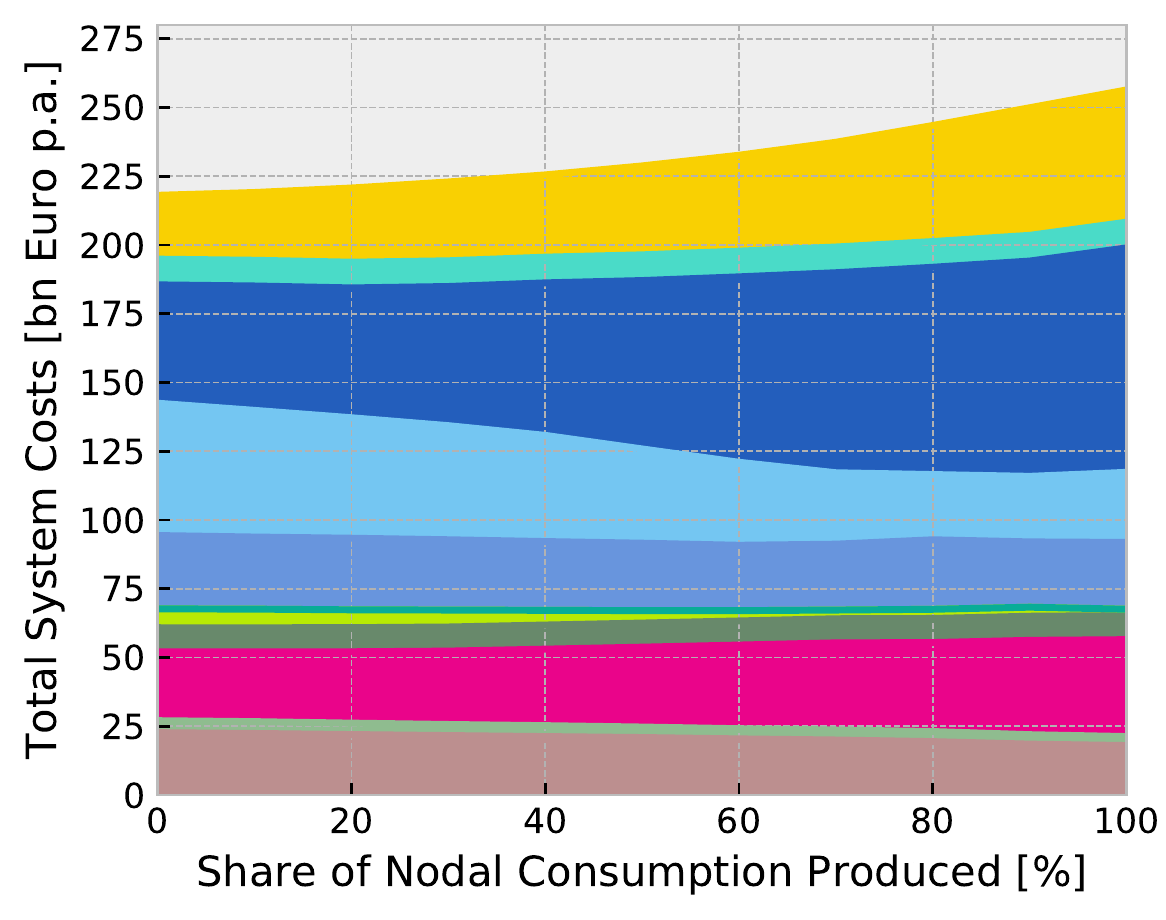}
	\includegraphics[width=0.15\textwidth, trim=12cm 1cm 0 0, clip]{{graphics/sensitivity-0.0}.pdf}
	\vspace{-0.35cm}
	\caption{Sensitivity of system cost and composition to nodal and country-wide equity requirements.}
	\label{fig:tsc}
	\vspace{-0.3cm}
\end{figure*}


Foremost, Figure \ref{fig:tsc} displays the sensitivity of system costs towards
nodal and country-wide equity requirements.
Similar graphics were produced regarding the amount of cross-border
transmission capacities by Schlachtberger et al.~\cite{schlachtberger_benefits_2017}.
National equity constraints cause a limited rise in total system cost.
The cost increase by less than 4\% when every country produces as much as they consume;
and by less than 2\% when each produces at least 80\%.
They entail less grid reinforcement and some more solar installations.
Conversely, the cost sensitivity is considerably higher for nodal equity constraints.
When every node on average produces all they consume, costs inflate by 18\%;
and already at equity levels of 50\% costs increase by 5\%.
Note, that the sensitivity is nonlinear.
Nodal requirements shift expansion plans towards onshore wind, solar and hydrogen storage,
while reducing network expansion and offshore wind capacities.
This confirms but also extends on a finding by Sasse et al.~\cite{sasse_distributional_2019}:
indeed solar contributes to regional equity, but also onshore wind does.
This is ambivalent since onshore wind is susceptible to local opposition.

\begin{figure}
	\begin{center}
		\begin{scriptsize}
			(i) Optimum (No Requirements)
			\includegraphics[width=.9\columnwidth, trim=0 5cm 0 0.5cm, clip]{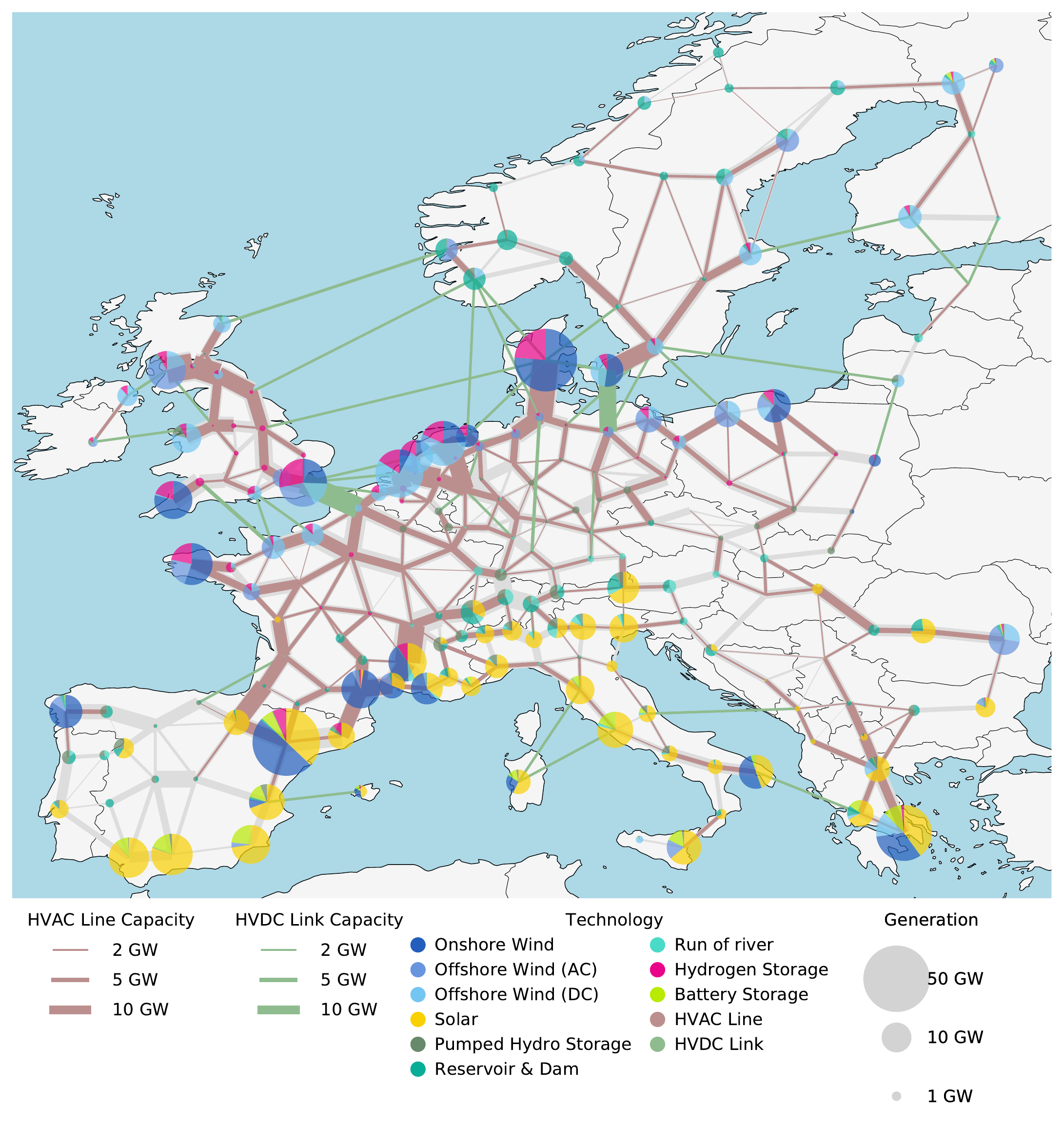} \\
			(ii) National Balance (100\%)
			\includegraphics[width=.9\columnwidth, trim=0 5cm 0 0.5cm, clip]{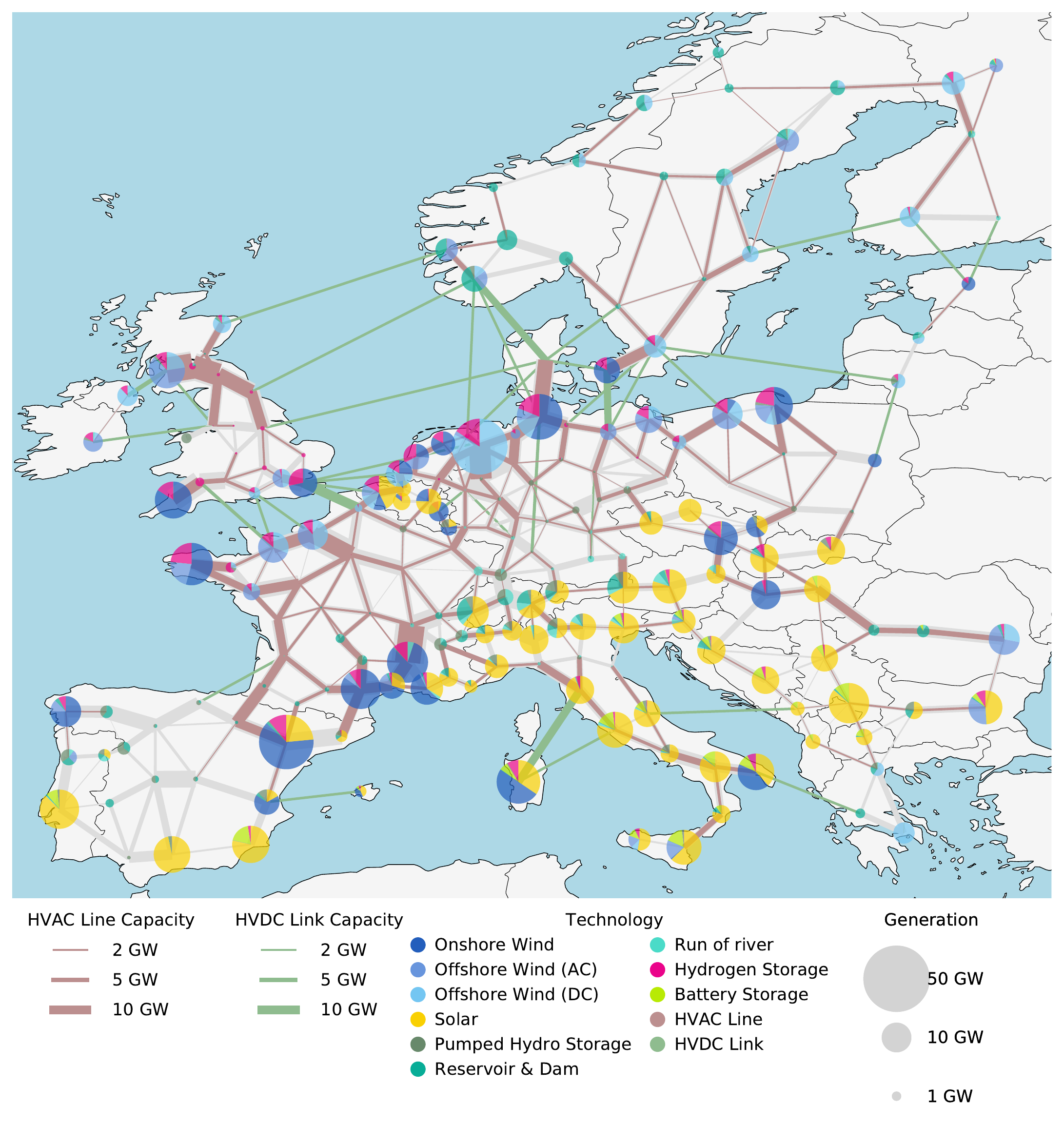} \\
			(iii) Nodal Balance (100\%)
			\includegraphics[width=.9\columnwidth, trim=0 0cm 0 0.5cm, clip]{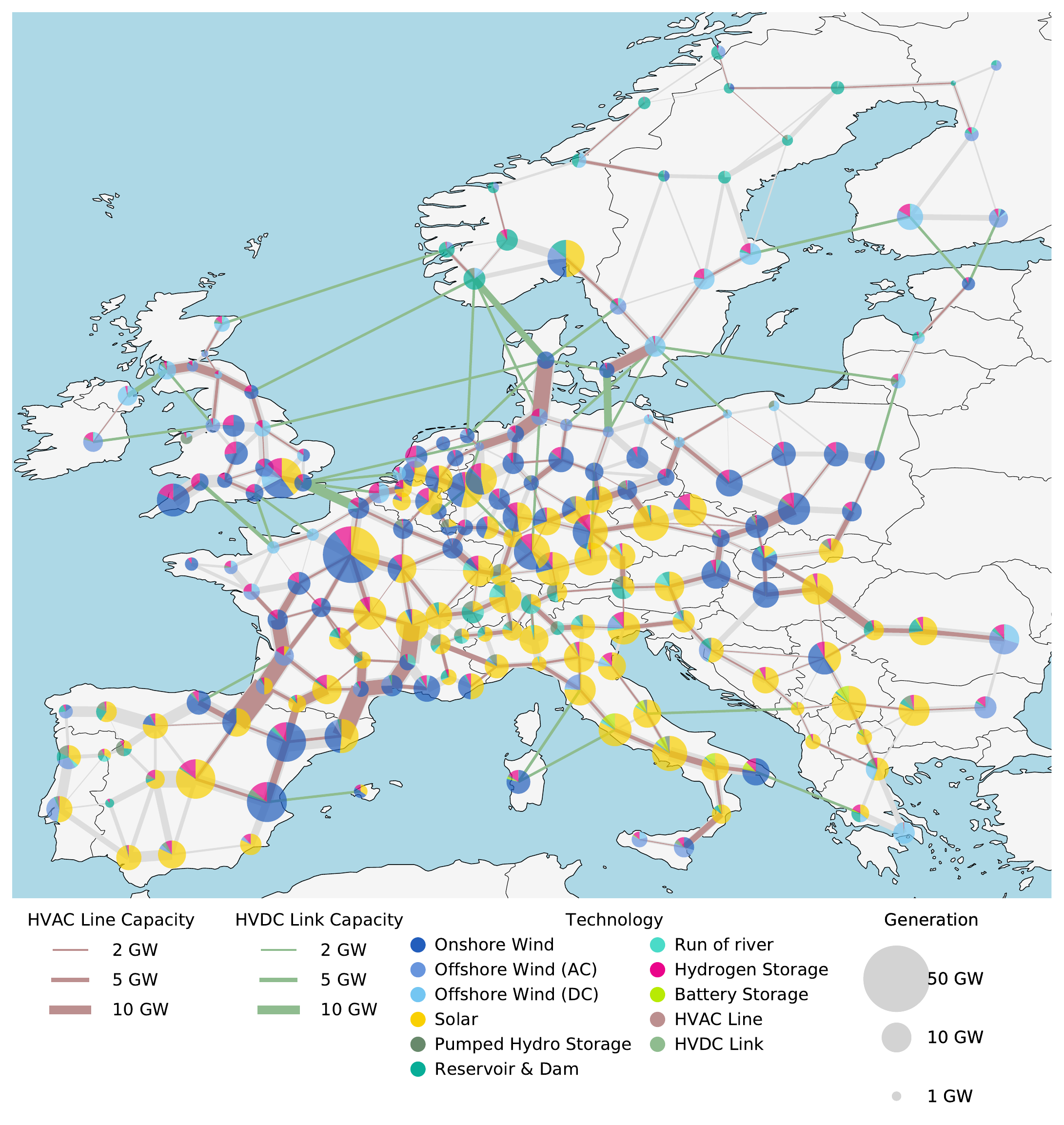} 
		\end{scriptsize}
	  \end{center}
	  \vspace{-0.5cm}
	\caption{Maps of optimal system capacities for different equity requirements.}
	\label{fig:map}
\end{figure}

The maps of optimal system capacities in Figure \ref{fig:map}
show less but still substantial amounts of transmission expansion
in the case of nodal equity.
Compared to the unrestricted optimal solution,
the deployment of solar panels progresses northbound
and onshore wind capacities spread in Northern and Eastern Europe.
Moreover, storage infrastructure distributes more evenly.

Figure \ref{fig:lorenz} depicts Lorenz curves as equity measures
for different equity constraints (cf.~\cite{sasse_distributional_2019}).
They relate the cumulative share of electricity generation
to the cumulative share of demand in the 200 regions of the model.
The Lorenz curve is on the identity line if annual sums of
generation and load are equal at each node.
While nodal equity requirements by definition lift the Lorentz curve,
national requirements maintains an unequal distribution of infrastructure
within each country.

According to Li et al.~\cite{li_regional_2016}, winners are those who receive
high per-capita investments. 
Results show that, when every country balances generation and load on average,
the national annual investments relative to demand are more evenly
distributed, ranging between 40 and 100 \euro{}/MWh (Figure \ref{fig:invest}).
Generation infrastructure dominates investments followed by storage and transmission.
The absence of extreme winners and losers is favourable.

While even nodal production equity raises
costs only to a limited extent below 20\%,
absolute autarky is significantly more costly
already on a national level.
Figure \ref{fig:autarky-cost} shows that
eliminating cross-border transmission capacities
(i.e.~no trade of power between countries)
adds costs beyond 40\%. Costs rise even more when 
each of the 200 regions is fully self-sufficient.
With an additional 85\%, costs almost
doubled compared to the least-cost solution.

One critical factor concerns the land use requirements
for solar in the nodal autarky scenario.
Assuming an even split between utility and rooftop PV,
results reveal that no region uses more than 3\%
of its area for utility PV and only one in ten regions uses more than 1\%.
It further needs to be noted that hydrogen storage in salt caverns, as
the cheaper alternative to steel tanks
if the geological conditions admit it, was neglected.
In future work, the autarky analysis should be repeated
for a fully sector-coupled energy system which includes
transport options for chemical energy carriers.

\section{Conclusion}
\label{sec:conclusion}

It is possible to strike a balance between cost-efficiency
and fair distribution of investments at little additional expense.
Aligning annual generation and consumption per country costs 
less than 4\% more; per node the costs increase by 18\%.
National balancing however retains inhomogenous distributions within the countries.
Finally, even when each node produces as much as they consume,
power is still extensively transmitted and regions are not self-sufficient.
True autarky solutions without power transmission are in relation 
substantially more expensive.
Knowledge about the observed degree of freedom is important
considering that more even investment per region can lead to
better political feasibility, quicker implementation, and higher social acceptance.

\vspace{-0.4cm}
\section*{Acknowledgements}
F.N. acknowledges funding from the Helmholtz Association under grant no. VH-NG-1352.
The responsibility for the contents lies solely with the author.
\doclicenseLongText
\doclicenseIcon

\vspace{-0.4cm}
\bibliography{equity.bib}

\end{document}